\documentclass[letter]{aa}

\usepackage{graphicx}
\usepackage{amsmath}
\usepackage{times}
\usepackage{xspace}
\usepackage{color}
\usepackage{breqn}
\usepackage[varg]{txfonts}
\usepackage{natbib}

\usepackage[breaklinks,colorlinks,citecolor=blue]{hyperref} 
\usepackage[all]{hypcap} 

\bibpunct{(}{)}{;}{a}{}{,}
\usepackage[normalem]{ulem}
\usepackage{float}
\usepackage{epsfig}
\usepackage{multirow}
\usepackage{epstopdf}
\usepackage{verbatim}
\usepackage{nicefrac}
\usepackage{url}
\usepackage{bm}
\usepackage{amssymb}
\usepackage{booktabs}
\usepackage{threeparttable}
\usepackage{amsbsy}
\usepackage{amsfonts}
\usepackage{pgfplots}
\usepackage{upgreek}
\usepackage{subcaption}
\usepackage{fontawesome}
\usepackage{xcolor}

\usepackage{tabularx}
\newcolumntype{R}{>{\raggedleft\arraybackslash}X}

\def\aap{A\&A}

\newcommand{\sjaddress}{\url{https://github.com/sjoudaki/CosmoLSS}\xspace}

\newcommand{\diracaddress}{\url{https://dirac.ac.uk}\xspace}

\newcommand{\be}{\begin{equation}}
\newcommand{\ee}{\end{equation}}
\newcommand{\bea}{\begin{eqnarray}}
\newcommand{\eea}{\end{eqnarray}}

\newcolumntype{P}[1]{>{\centering\arraybackslash}p{#1}}

\newcommand{\halofit}{\textsc{halofit}\xspace}
\newcommand{\hmcode}{\textsc{hmcode}\xspace}

\newcommand{\bpz}{\textsc{bpz}\xspace}
\newcommand{\cosmomc}{\textsc{CosmoMC}\xspace}
\newcommand{\cosmolss}{\textsc{CosmoLSS}\xspace}
\newcommand{\cosmosis}{\textsc{CosmoSIS}\xspace}
\newcommand{\montepython}{\textsc{Monte Python}}

\newcommand{\camb}{\textsc{CAMB}\xspace}

\newcolumntype{x}{>{\raggedright\arraybackslash}X}

\begin{document}
\title{KiDS+VIKING-450 and DES-Y1 combined:\\ Cosmology with cosmic shear}
\titlerunning{Combined analysis of KV450 and DES-Y1}
\author
{S.~Joudaki\inst{1}
\and
H.~Hildebrandt\inst{2,3}
\and
D.~Traykova\inst{1}
\and
N.~E.~Chisari\inst{1}
\and
C.~Heymans\inst{4,2}
\and
A.~Kannawadi\inst{5}
\and
K.~Kuijken\inst{5}
\and
A.~H.~Wright\inst{2,3}
\and
M.~Asgari\inst{4}
\and
T.~Erben\inst{3}
\and
H.~Hoekstra\inst{5}
\and
B.~Joachimi\inst{6}
\and
L.~Miller\inst{1}
\and
T.~Tr\"{o}ster\inst{4}
\and
J.~L.~van~den~Busch\inst{2,3}
}

\authorrunning{The KiDS Collaboration}

\institute{
Department of Physics, University of Oxford, Denys Wilkinson Building, Keble Road, Oxford OX1 3RH, UK
\and
Ruhr-Universit\"at Bochum, Astronomisches Institut, German Centre for Cosmological Lensing, Universit\"atsstr. 150, 44801, Bochum, Germany
\and
Argelander-Institut f\"ur Astronomie, Universit\"at Bonn, Auf dem H\"ugel 71, 53121 Bonn, Germany
\and
Institute for Astronomy, University of Edinburgh, Royal Observatory, Blackford Hill, Edinburgh EH9 3HJ, UK
\and
Leiden Observatory, Leiden University, P.O.~Box 9513, 2300 RA Leiden, The Netherlands
\and
Department of Physics and Astronomy, University College London, Gower Street, London WC1E 6BT, UK
}

\date{Received June 21, 2019; accepted April 22, 2020}

\abstract{
We present a combined tomographic weak gravitational lensing analysis of the Kilo Degree Survey (KV450) and the Dark Energy Survey (DES-Y1). We homogenize the analysis of these two public cosmic shear datasets by adopting consistent priors and modeling of nonlinear scales, and determine new redshift distributions for DES-Y1 based on deep public spectroscopic surveys. Adopting these revised redshifts results in a $0.8\sigma$ reduction in the DES-inferred value for $S_8$, which decreases to a $0.5\sigma$ reduction when including a systematic redshift calibration error model from mock DES data based on the MICE2 simulation. The combined KV450 + DES-Y1 constraint on $S_8 = 0.762^{+0.025}_{-0.024}$ is in tension with the Planck 2018 constraint from the cosmic microwave background at the level of $2.5\sigma$. This result highlights the importance of developing methods to provide accurate redshift calibration for current and future weak lensing surveys.
}
\maketitle

\defcitealias{Hildebrandt19}{H20}
\defcitealias{Troxel18}{T18}

\section{Introduction}
\label{introsec}
\setcounter{footnote}{0}
\renewcommand{\thefootnote}{\arabic{footnote}}

Weak gravitational lensing tomography has entered the phase of precision cosmology, with observational constraints on the best-measured parameter, $\smash{S_8 = \sigma_8 \sqrt{\Omega_{\rm m}/0.3}}$, at a level of precision $\lesssim5\%$ for all current surveys (\citealt{Hildebrandt19}, hereafter \citetalias{Hildebrandt19};~\citealt{Troxel18}, hereafter \citetalias{Troxel18};~\citealt{Hikage19,joudaki16,jee2016}). Here, $\sigma_8$ refers to the root-mean-square of the linear matter overdensity field on $8 \, h^{-1} \, {\rm Mpc}$ scales, and $\Omega_{\rm m}$ is the present mean density of non-relativistic matter relative to the critical density. This phase has been reached as a result of the success in accounting for the systematic uncertainties that affect the measurements. However, as the statistical precision of weak lensing surveys increases with depth and area, the requirements on their ability to control systematic uncertainties increase as well. In \citet{Hildebrandt16}, it was shown that the contribution of systematic uncertainties to the total error budget for the Kilo Degree Survey (KiDS;~\citealt{kuijken15}) is comparable to that of the statistical uncertainties. Given the similar constraining power of concurrent weak lensing surveys, such as the Dark Energy Survey (DES;~\citealt{desdr1}) and the Subaru Hyper Suprime-Cam survey (HSC;~\citealt{aihara18a}), a continued reduction in the systematic uncertainties is crucial to obtain unbiased cosmological constraints and to exploit the full statistical power of current and future weak lensing datasets.

The most notable systematic uncertainties pertain to the intrinsic alignment (IA) of galaxies, additive and multiplicative shear calibration, baryonic feedback affecting the nonlinear matter power spectrum, and photometric redshift errors (see \citealt{mandelbaum18} and references therein). All current weak lensing surveys have reached a statistical precision where notable changes to the cosmological parameter constraints are found when accounting for these systematic uncertainties in the analysis (e.g.~\citealt{Hikage19}; \citetalias{Troxel18,Hildebrandt19}). The expectation is that the final parameter constraints are robust when marginalized over all known systematics. This is generally well-motivated through the vast range of checks and extensions of the systematic models beyond the standard approach considered by these surveys. The uncertainty in the redshift distributions, $n(z)$, of weakly lensed galaxies is, however, more difficult to account for, and has been shown to be the only systematic uncertainty to impact the posterior mean of $S_8$ by $\sim1\sigma$ \citepalias{Hildebrandt19}.

The redshift uncertainty is arguably the most challenging systematic to control in both current and future lensing surveys. In KiDS, the estimation of the redshift distributions has benefited from the fully overlapping near-infrared imaging data from the VISTA Kilo-Degree Infrared Galaxy Survey (VIKING;~\citealt{viking13}). The combined KiDS and VIKING dataset (`KiDS+VIKING-450' or `KV450';~\citealt{Wright19}) has allowed for an increased precision in the estimation of photometric redshifts that are used to assign sources to tomographic bins. In addition, KiDS targets deep pencil-beam spectroscopic surveys permitting the redshift distributions to be determined via the weighted direct estimation, or `DIR', approach (\citealt{lima08,Hildebrandt16}; \citetalias{Hildebrandt19}), which is fully decoupled from the photo-$z$. This DIR method assigns KiDS sources to spectroscopic galaxies via a $k$-nearest-neighbour matching in order to estimate weights for the spectroscopic objects. The weighted distribution of spectroscopic redshifts can then be used to estimate the $n(z)$ of the sources. The uncertainty $\Delta z_i$ in the mean redshift of each tomographic bin $i$ is obtained from a spatial bootstrap resampling of the spectroscopic calibration sample and propagated in the cosmological analysis as $n_i(z) \rightarrow n_i(z-\Delta z_i)$ \citepalias{Hildebrandt19}.

The DIR approach has been found to produce cosmological results consistent with other $n(z)$ estimation techniques, such as the angular cross-correlation of photometric and spectroscopic galaxy samples (where the spectroscopic samples are obtained from overlapping wide and shallow surveys;~\citealt{morrison17,johnson17}). In \citetalias{Hildebrandt19}, it was also shown that the cosmological constraints from KV450 are robust to the specific combination of spectroscopic calibration samples used to obtain the DIR $n(z)$ as long as the spectroscopic datasets provide a sufficient coverage in depth and redshift.

Both DES and HSC calibrate their redshift distributions with a high-quality photometric redshift catalogue in the COSMOS field \citep{laigle16}. A similar calibration of the KV450 data yielded a $0.6\sigma$ larger value of $S_8$ \citepalias{Hildebrandt19}. One hypothesis is that outliers in the COSMOS photo-$z$ catalogue cause the estimated redshifts to be biased low. Alternatively, there could be a bias in the fiducial KV450 DIR calibration. Here, we construct mock KV450 and DES-Y1 catalogues based on the MICE2 simulation and quantify the extent to which the redshift distributions might be reliably estimated. As the DES-Y1 data are slightly shallower than KiDS, which matches the depth of the public spectroscopic redshift catalogues, we spectroscopically calibrate the DES-Y1 redshift distributions.\footnote{The HSC-Y1 shear catalogues were not publicly released at the time of this work, and their greater depth also makes a direct spectroscopic calibration infeasible.} Using these newly determined $n(z)$, we evaluate the impact on the cosmological constraints, and perform a combined cosmological analysis with KV450.

\section{KV450 and DES-Y1 cosmological constraints with a homogenized analysis}
\label{cosmosec1}

\begin{figure}
\vspace{-1em}
\begin{center}
\resizebox{9.6cm}{!}
{\includegraphics{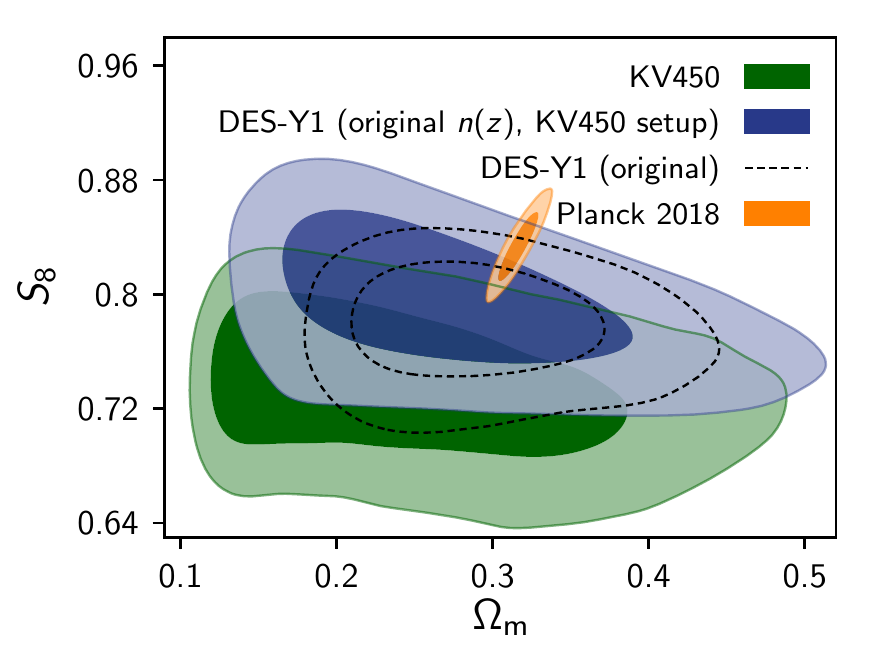}}
\end{center}
\vspace{-1.9em}
\caption{
Marginalized posterior contours in the $S_8$ -- $\Omega_{\mathrm m}$ plane (inner 68\%~CL, outer 95\%~CL).
We show the KV450 constraints in green (solid) using an analysis setup that follows \citetalias{Hildebrandt19}, but including an additional redshift dependence of the IA signal (denoted `KV450'). In black (dashed), we show the DES-Y1 constraints corresponding to the original \citetalias{Troxel18} analysis, noting that the sum of neutrino masses is varied in this analysis (and hence the contour should not be directly compared with the orange (solid) Planck 2018 contour where neutrino mass is fixed). The blue (solid) contours show the DES-Y1 constraints where an identical setup to the KV450 analysis is used (along with the original DES-Y1 redshift distributions).
}
\label{fig1label}
\end{figure}

To meaningfully compare the cosmological constraints from KV450 and DES-Y1, we begin by homogenizing the cosmological priors and treatment of astrophysical systematic uncertainties (Fig.~\ref{fig1label}). We consider the KV450 and DES-Y1 measurements and covariance in \citetalias{Hildebrandt19} and \citetalias{Troxel18}, respectively.\footnote{A unified analysis of earlier cosmic shear datasets is performed in \citet{chang19}.} We do not remeasure the respective data vectors and covariance, and use only the angular scales advocated in \citetalias{Hildebrandt19} and \citetalias{Troxel18}. As KV450 and DES-Y1 observations do not overlap on the sky, we treat the two surveys as distinct.

The cosmological constraints on KV450 and DES-Y1 are obtained using the \cosmolss\footnote{\sjaddress} \href{https://github.com/sjoudaki/CosmoLSS}{\faGithub} likelihood code \citep{joudaki18} in a Markov Chain Monte Carlo (MCMC) analysis. This code has been used to benchmark the LSST-DESC Core Cosmology Library's (CCL;~\citealt{chisari19}) computation of tomographic cosmic shear, galaxy-galaxy lensing, and galaxy clustering observables. For completeness, we reproduced the \cosmolss DES-Y1 constraints with both \cosmosis \citep{cosmosis} and the Planck Collaboration's lensing likelihood in \cosmomc \citep{planck18}. In \citetalias{Hildebrandt19}, we moreover showed that the KV450 constraints from \cosmolss, \cosmosis, and \montepython~\citep{montepython13} are in excellent agreement.

For both surveys, we implement the cosmological priors of \citetalias{Hildebrandt19} (see Table 3 therein). In the case of DES-Y1, this includes not only a change in the size of the parameter priors, but notably also a change in the size of the parameter space by fixing the sum of neutrino masses to 0.06 eV instead of varying it freely, a change in the uniform sampling of $A_{\mathrm s} \rightarrow \ln (10^{10} A_{\mathrm s})$, and a change from \halofit \citep{Takahashi12} to \hmcode \citep{Mead15} for the modeling of the nonlinear corrections to the matter power spectrum. Compared to the fiducial DES-Y1 and KV450 analyses, we also switch from \textsc{Multinest} \citep{mnest2} to MCMC sampling of the parameter space. Following \citetalias{Hildebrandt19}, we allow baryonic feedback to modify the nonlinear matter power spectrum. This does not particularly affect the DES-Y1 constraints given the conservative scale cuts in \citetalias{Troxel18}. We keep the shear calibration and photometric redshift uncertainties distinct between the two surveys (given by Table 2 in \citetalias{Troxel18} and Table 3 in \citetalias{Hildebrandt19}, respectively).

\begin{figure*}
\includegraphics[width=\hsize]{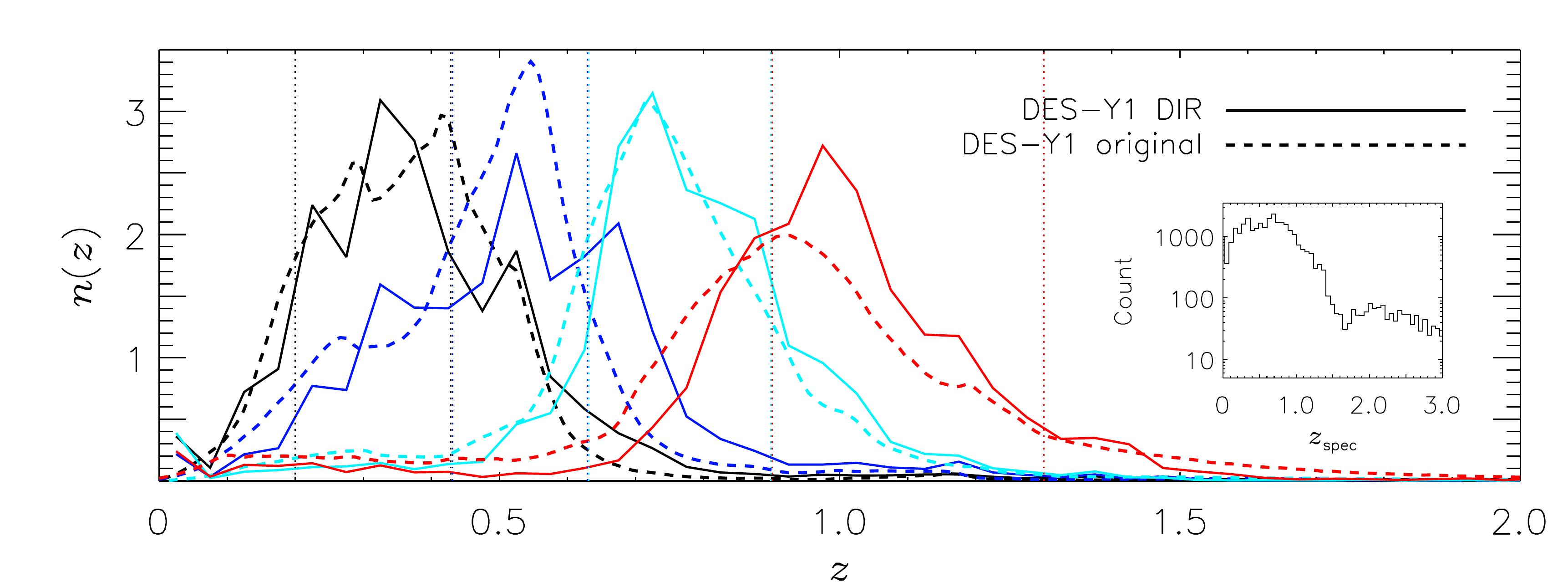}
\caption{DES-Y1 redshift distributions for the four tomographic bins (in black, blue, cyan, red, respectively), showing the publicly released distributions (dashed) and the spectroscopically determined distributions using the DIR approach (solid). The distributions based on spectroscopy are systematically shifted to larger redshifts compared to the original distributions (accounting for $\Delta z_i$), and hence favor a lower value of $S_8$ compared to the original DES-Y1 analysis in \citetalias{Troxel18}. See Table~\ref{tab1label} for the mean redshifts of the different tomographic bins for the two approaches. The vertical dotted lines denote the tomographic bin boundaries. The small inset shows the redshift distribution of the matched photometry/spectroscopy catalogue for DES-Y1 containing approximately $30,000$ objects used in the DIR method. The spectroscopic calibration samples are obtained from zCOSMOS, VVDS-Deep (2h), CDFS, DEEP2 (2h), VVDS-Wide (22h). We do not show the uncertainties in the $n(z)$ for visual clarity (instead see Table~\ref{tab1label} for uncertainties in the mean redshifts).
}
\label{fig2label}
\end{figure*}

Conservatively, we allow KV450 and DES-Y1 to have independent parameters governing the IA, using both an amplitude and redshift dependence (as a result, in the combined KV450 + DES-Y1 analysis there are 4 free IA parameters). We use a pivot redshift of $z_0 = 0.3$, in agreement with past KiDS analyses and direct measurements of the IA (e.g.~\citealt{mandelbaum11,joachimi11}). We find that the $S_8$ constraints are robust to the specific treatment of the IA, such as removal of the redshift dependence or by assuming that the IA parameters are shared between the two surveys.\footnote{We note, however, that a widened prior on $\eta_{\rm IA}$ allows for an extended confidence interval at low $\{\Omega_{\rm m}, S_8\}$ for DES-Y1 alone.}

We compare the KV450 and DES-Y1 constraints with the Planck 2018 cosmic microwave background (CMB) temperature and polarization measurements \citep{planck18},\footnote{Our comparisons are against the public chains, as the Planck 2018 likelihoods were not publicly released at the time of this work. This is not fully self-consistent given the mostly narrower prior ranges used by Planck (compared to our KV450 and DES-Y1 runs), but has a negligible impact given the constraining power of the Planck dataset.} where the `TT,TE,EE+lowE' data combination gives $S_8 = 0.834^{+0.016}_{-0.016}$. We exclude the CMB lensing measurements to isolate the high-redshift CMB temperature and polarization constraint on cosmology from the low-redshift Universe.

The KV450 constraint on $S_8 = 0.735^{+0.042}_{-0.034}$ corresponds to a $2.4\sigma$ discrepancy with Planck 2018. The original DES-Y1 cosmic shear constraint from the publicly released chain\footnote{\url{http://desdr-server.ncsa.illinois.edu/despublic/y1a1_files/chains/s_l3.txt}} is $S_8 = 0.778^{+0.030}_{-0.023}$ (we note that \citetalias{Troxel18} quotes the marginal posterior maximum of 0.782 instead of the more common posterior mean given here). Compared with the corresponding Planck 2018 result, where the neutrino mass varies, this is a $1.7\sigma$ difference. The DES-Y1 constraint using the KV450 setup is $S_8 = 0.794^{+0.037}_{-0.034}$, which differs by $1.0\sigma$ from the Planck 2018 constraint and by $1.1\sigma$ from the KV450 constraint. This change reflects a shift in the posterior mean and an increase in uncertainty as a result of using \hmcode instead of \halofit, wider priors on the amplitude and spectral index of the primordial power spectrum, uniformly sampling $\ln (10^{10} A_{\mathrm s})$ instead of $A_{\mathrm s}$, and fixing the sum of neutrino masses instead of varying it.

We note that when KV450 and DES-Y1 are homogenized to the same assumptions and using the fiducial angular scales, the constraining power of the two datasets is comparable, with the DES-Y1 uncertainty in $S_8$ smaller by 8\% (instead of 30\% smaller uncertainty when simply comparing the DES-Y1 constraint in \citetalias{Troxel18} with the KV450 constraint in \citetalias{Hildebrandt19}). However, this does not account for the improvement in the DES-Y1 constraining power when extending the scale cuts from the fiducial approach in \citetalias{Troxel18} to better agree with the range of angular scales $\theta$ probed by KV450. We find that such a modification to the angular scales (such that $\{\theta_+ > 3, \theta_- > 7\}~{\rm arcmin}$ for all tomographic bin combinations) in our correlation function analysis improves the DES-Y1 uncertainty in $S_8$ by approximately 30\% (with a $0.5\sigma$ decrease in the posterior mean) after marginalizing over baryonic feedback, increasing the deviation from Planck (see also \citealt{asgari20} for a small-scale analysis with COSEBIs).

\section{Spectroscopic determination of the DES-Y1 source redshift distributions}
\label{specsec}

The redshift distributions for DES and HSC have so far been obtained by using data from the 30-band photometric dataset `COSMOS-2015' \citep{laigle16}. In HSC-Y1, the fiducial redshift distributions are obtained as a histogram of reweighted COSMOS-2015 photometric redshifts (using the weights of the HSC source galaxies and a self-organizing map, or `SOM'), and the uncertainties in these distributions are obtained by comparing against the \emph{photometric} redshift distributions from six different codes where the probability distribution functions of the source galaxy redshifts are stacked \citep{Hikage19}. In DES-Y1, the Bayesian photometric redshift code \bpz \citep{benitez2000} is used to compute a stacked redshift distribution, which is shifted along the redshift axis to best fit a combination of resampled COSMOS-2015 redshift distributions and (for the first three tomographic bins) the clustering of the DES source galaxies and a high-quality photo-$z$ reference sample (\textsc{redMaGiC};~\citealt{Rozo16}) over a limited redshift range \citep{Hoyle18}.

\begin{table}
\setlength\tabcolsep{18pt}
  \caption{\label{tab1label} DES-Y1 mean redshifts of the four tomographic bins calibrated with COSMOS-2015 \citepalias{Troxel18} and spectroscopic redshifts (this work). The spectroscopic calibration consistently favors distributions with larger mean redshifts compared to COSMOS-2015 (the same is found for the median redshifts). We note that our mock analysis of the spectroscopic calibration based on the MICE2 simulation suggests lower mean redshifts by approximately $0.01$--$0.03$ depending on the tomographic bin (see Appendix~\ref{appmice} for details on the mock analysis along with a discussion of its limitations and cosmological implications).
  }
  \begin{tabular}{ccc}
    \hline
    \hline
    Tom. & {COSMOS-2015} & {Spec-$z$ (DIR)}\\
          bin  & $<z>$ & $<z>$ \\
    \hline
    1 & $0.389 \pm 0.016$  & $0.403 \pm 0.008$\\
    2 & $0.507 \pm 0.013$  & $0.560 \pm 0.014$\\
    3 & $0.753 \pm 0.011$  & $0.773 \pm 0.011$\\
    4 & $0.949 \pm 0.022$  & $0.984 \pm 0.009$\\
    \hline
  \end{tabular}
\end{table}

To compare these approaches to direct spectroscopic determination, which fully decouples the photo-$z$ from the determination of the $n(z)$, \citetalias{Hildebrandt19} considered a DIR estimate of the KV450 redshifts with the help of COSMOS-2015, finding a coherent downward shift in the redshift distributions and a consequent increase in the posterior mean for $S_8$. \citetalias{Hildebrandt19} argue that estimating the redshift distributions from COSMOS-2015 might however be unreliable given the `catastrophic outlier' fraction of $\sim$6\% in the magnitude range $23 < i < 24$ reported in \citet{laigle16}\footnote{For $22 < i < 23$, the outlier rate is significant at 3.5\% (O. Ilbert, private communication).} and a residual photo-$z$ bias of $\langle z_{\mathrm{spec}} - z_{\mathrm{phot}} \rangle \approx 0.01$ after rejection of outliers. This can be compared to $\sim$1\% unreliable redshifts for the combined spectroscopic calibration sample.\footnote{In \citet{wright20}, we show that the change in the estimated redshift distributions from catastrophic spec-$z$ failures in the spectroscopic compilation is negligible.} The outliers in the COSMOS-2015 photo-$z$ are potentially also more problematic because their effect is most probably asymmetric. Outliers that are truly objects at high-$z$ but are assigned a low COSMOS-2015 photo-$z$ are more likely to fall inside the DES-Y1 tomographic bins than outliers that are bona-fide low-$z$ galaxies but are assigned a high COSMOS-2015 photo-$z$. Additionally, the bias in the core of the $z_{\mathrm{spec}} - z_{\mathrm{phot}}$ distribution is in the same direction, i.e. overall the redshifts might be underestimated by the COSMOS-2015 photo-$z$.

In the DES-Y1 analyses, the case is made that a spectroscopic determination of the source redshift distributions would not be sufficiently accurate due to the incompleteness of the existing spectroscopic surveys at the faint end of the DES observations \citep{Hoyle18}. We find, however, that even the deeper KV450 source sample is well covered by our spectroscopic compilation, implying that the coverage should also be sufficient for the calibration of the DES-Y1 sample. This is confirmed by a SOM approach to redshift calibration \citep{Masters15} presented in \citet{wright20}. 

DES-Y1 overlaps with almost the same deep spectroscopic redshift surveys that were used by \citetalias{Hildebrandt19}. As shown in Fig.~\ref{fig2label} (inset), this overlap contains some 30,000 objects with spectroscopic redshifts from zCOSMOS \citep{zcosmos09}, the DEEP2 Redshift Survey \citep{deep13}, the VIMOS VLT Deep Survey (VVDS;~\citealt{vvds13}), and the Chandra Deep Field South (CDFS;~\citealt{vanzella08,popesso09,balestra10,vvds13}). We find that the KV450 source sample is well covered as long as spectroscopic redshifts from DEEP2 -- the highest-redshift calibration survey -- are included and the same is true for DES-Y1. However, we note that \citet{Hoyle18} and \citet{hartley20} have moreover argued that the 4-band DES data may be inherently less suitable to our re-weighting scheme than the 9-band KiDS+VIKING data, which is a hypothesis that we assess in Appendix~\ref{appmice} (see also \citealt{buchs19} for a way to solve this by leveraging the DES Deep fields).

The KV450 and DES-Y1 spectroscopic calibration samples used here differ in detail: DES-Y1 overlaps on the sky with VVDS in both the Deep (2h) and Wide (22h) fields compared to only the Deep (2h) field for KV450, and the DES-Y1 calibration does not include the 23h field of DEEP2 and the GAMA-G15Deep sample \citep{kafle18} which are included in the KV450 calibration. Overall, we obtain the DES-Y1 and KV450 redshift distributions using five and six spectroscopic calibration samples, respectively, of which four are identical.\footnote{Note that the exact area in each of these fields differs slightly between surveys because of the different footprints of KiDS and DES.} Note that no shear data from these calibration fields are used in both the KiDS and DES cosmological analyses, maintaining independence in the measured shear correlation functions from the two surveys.

\begin{figure}
\vspace{-1em}
\begin{center}
\resizebox{9.6cm}{!}
{\includegraphics{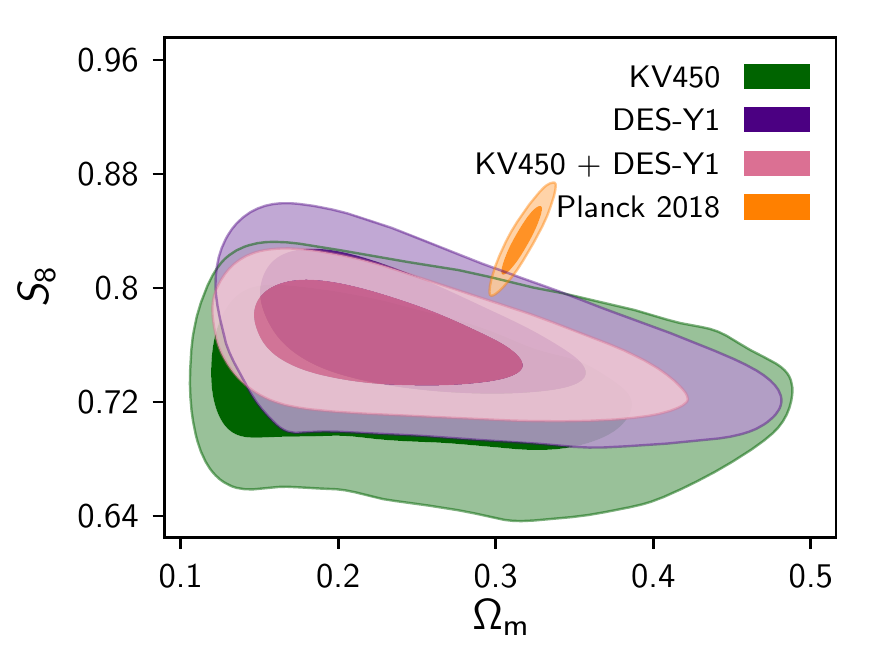}}
\end{center}
\vspace{-1.9em}
\caption{
Marginalized posterior contours in the $S_8$ -- $\Omega_{\mathrm m}$ plane (inner 68\%~CL, outer 95\%~CL) for KV450 (green), DES-Y1 following a spectroscopic calibration of the redshift distributions and identical setup to the KV450 analysis (purple), the above combined (pink), and Planck 2018 (orange). 
}
\label{fig3label}
\end{figure}

Figure~\ref{fig2label} shows that the spectroscopic calibration shifts DES-Y1 redshift distributions to higher redshifts compared to the original photo-$z$ recalibration with COSMOS-2015, consistent with the findings of \citetalias{Hildebrandt19}. Mean redshifts of the four tomographic bins are reported in Table~\ref{tab1label} for both cases. The spectroscopically determined distributions peak closer to the centre of the corresponding tomographic bins, and contain higher-redshift galaxies. These shifts between the spectroscopically estimated and published DES-Y1 $n(z)$ are significant because of their coherence, i.e. all tomographic bins shift in the same direction. We emphasize that widening the priors on the uncorrelated $\Delta z_i$ nuisance parameters cannot account for such a coherent shift as this is fully degenerate with the cosmological parameters of interest (see the discussion at the end of section 3 in \citetalias{Hildebrandt19}).

In Appendix~\ref{appmice}, we further explore the robustness of the DIR calibration on mock KV450 and DES-Y1 catalogues. This analysis motivates the inclusion of a systematic error model in our analysis to account for potential biases in the DIR calibration. If the error model from the mock survey analysis is fully accurate and representative (see the caveats and discussion in Appendix~\ref{appmice}), the true mean redshifts of the DES-Y1 tomographic bins can be lowered by approximately $0.01$--$0.03$ compared to the DIR results presented in Fig.~\ref{fig2label} and Table~\ref{tab1label}.

\section{Cosmological impact of DES-Y1 $n(z)$ recalibration and combined constraints with KV450}
\label{cosmosec2}

We now quantify the impact of the spectroscopic calibration of the DES-Y1 redshift distributions on the cosmological parameter constraints. As it is now on an equal footing with KV450, we moreover perform a combined analysis of the two surveys, shown in Fig.~\ref{fig3label}.

The DES-Y1 constraint following the spectroscopic calibration of the redshift distributions is $S_8 = 0.765^{+0.036}_{-0.031}$. Compared to using the original redshift distributions, this is a change in the posterior mean by $\Delta S_8 = -0.029$ and a marginal ($5\%$) improvement in the $S_8$ uncertainty. We verified that this shift in $S_8$ is largely recovered by coherently shifting the original DES-Y1 redshift distributions by the $\Delta z_i$ difference with the spectroscopically calibrated distributions as reported in Table~\ref{tab1label} (i.e.~changes in the structure of the $n_i(z)$ have a subdominant impact on $S_8$). This substantial change in the DES-Y1 constraint highlights the importance of the redshift calibration. The size of $\Delta S_8$ corresponds to a $0.8\sigma$ shift in terms of the larger DES uncertainty in the KV450 setup, and a $1.1\sigma$ shift in terms of the original DES-Y1 uncertainty quoted in \citetalias{Troxel18}. The DES-Y1 constraint using a KV450 analysis setup and spectroscopically calibrated redshift distributions is different from the Planck 2018 constraint on $S_8$ by $1.9\sigma$. The goodness of fit with the spectroscopically calibrated distributions is comparable to that of using the COSMOS-2015 distributions (difference in the reduced $\chi^2$ by $6\times10^{-3}$). 
  
Following the homogenization of the analysis setups, the combined KV450 + DES-Y1 constraint is $S_8 = 0.762^{+0.025}_{-0.024}$. This is almost exactly a factor of $\sqrt{2}$ improvement in precision compared to KV450 and DES-Y1 on their own. We find a best-fit $\chi^2 = 413.4$ for 397 degrees of freedom, which corresponds to a reduced $\chi^2$ of 1.04 and a $p$-value of 0.27. Using the $\log \mathcal{I}$ statistic \citep{joudaki16} and Jeffreys' scale \citep{jeffreys,kr95}, we find that KV450 and DES-Y1 are in `strong' concordance ($\log \mathcal{I} = 1.4$), which is an expected outcome given the $S_8$ agreement between the two surveys. The KV450 + DES-Y1 constraint is $2.5\sigma$ discordant with Planck 2018 (we do not evaluate the $\log \mathcal{I}$ statistic in this case as the Planck 2018 likelihood is not public). We note that for the cosmological priors used in \citetalias{Troxel18}, the combined KV450 + DES-Y1 dataset is even more discordant with Planck. For this case (not shown in Fig.~\ref{fig3label}), $S_8 = 0.750^{+0.022}_{-0.025}$, which is a $3.0\sigma$ discordance with Planck 2018. 

The constraints on the astrophysical degrees of freedom, such as the IA amplitude and redshift dependence, do not change significantly in the combined analysis from either survey independently. This is partly a consequence of our analysis decision to keep the KV450 and DES-Y1 intrinsic alignment parameters distinct. We further note that the impact of the spectroscopic calibration for DES-Y1 decreases to $\Delta S_8 = -0.017$ (from the fiducial $\Delta S_8 = -0.029$) if a systematic error model for the DIR calibration from our study of mock DES data in Appendix~\ref{appmice} is included in the analysis. In the appendix, we show that a self-consistent change in the redshift distributions of both DES-Y1 and KV450, based on our mocks constructed for each survey, results in effectively the same combined KV450 + DES-Y1 constraint on $S_8$ as in the fiducial analysis (less than $0.1\sigma$ difference). While the inclusion of the DEEP2 sample is critical for the redshift calibration of both KV450 and DES-Y1 \citep{wright20}, the $S_8$ constraints from both surveys are robust to a change in the spec-$z$ calibrating fields to the four fields that they have in common. We note that the spectroscopically calibrated source redshift distributions will have a comparable impact on the $S_8$ constraint from the DES-Y1 combined analysis of cosmic shear, galaxy-galaxy lensing, and galaxy clustering \citep{desy1main}.

\section{Conclusions}
We have performed the first combined analysis of Stage-III cosmic shear surveys with KiDS+VIKING-450 and DES-Y1. In obtaining reliable cosmological results, we homogenized the analysis setups and spectroscopically calibrated the DES-Y1 source redshift distributions, both of which have a substantial impact on the parameter constraints. We show that the cosmological constraints from KV450 and DES-Y1 are comparable when analyzed self-consistently over the angular scales advocated by each survey, and that the DES-Y1 constraint on $S_8$ changes downwards by $0.8\sigma$ when calibrating the redshift distributions using overlapping deep-field spectroscopy. The combined KV450 + DES-Y1 constraint on $\smash{S_8 = 0.762^{+0.025}_{-0.024}}$ reflects a factor of $\sqrt{2}$ improvement in precision compared to each survey independently, and is $2.5\sigma$ discordant with the Planck CMB temperature and polarization. This increases to $3.0\sigma$ when employing the cosmological priors advocated by DES-Y1, and would only increase further by including smaller-scale DES-Y1 measurements sensitive to baryonic feedback.

The substantial change in the DES-Y1 redshift distributions and the corresponding impact on the $S_8$ constraint suggests that a similar exercise with HSC-Y1 data would be valuable, and that a self-consistent combined analysis of all three current cosmic shear surveys may sharpen the tension with Planck 2018 even further. We note that the greater depth of HSC (but also future surveys such as LSST) complicates a direct spectroscopic calibration of the redshift distributions and may instead require other approaches such as the cross-correlation between photometric and spectroscopic galaxies \citep{newman08}. Ultimately, the advent of additional data expected for KiDS, DES, and HSC in the coming years along with self-consistent combined analyses of cosmic shear surveys will be crucial to resolving the current tension found with the Planck CMB.

\section*{Acknowledgements}
We thank Chris Blake, Pedro Ferreira, Christos Georgiou, Ian Harrison, Olivier Ilbert, Harry Johnston, Nicolas Martinet, Alexander Mead, Chris Morrison, and Mohammadjavad Vakili for useful discussions. We also thank Ian Harrison for help navigating the public DES data. We thank the DES team and in particular Daniel Gr\"un and Michael Troxel for in-depth discussions that led to the inclusion of the simulation results reported in the appendix. We acknowledge the use of \camb/\cosmomc packages (\citealt*{LCL}; \citealt{Lewis:2002ah}).
\newline
{\it Author contributions:} All authors contributed to the development and writing of this paper. The authorship list is given in three groups: the lead authors (SJ, HHi, DT), followed by two alphabetical groups. The first alphabetical group includes those who are key contributors to both the scientific analysis and the data products. The second group covers those who have either made a significant contribution to the data products or to the scientific analysis.

\bibliographystyle{aa}
\bibliography{kidsdes}

\begin{appendix}

\section{Tests on MICE2 mock catalogues}
\label{appmice}

We test the spectroscopic DIR calibration described in Sect.~\ref{specsec} on mock catalogues created from the public MICE2 simulation \citep{fosalba15,crocce15}. These mock catalogues are similar to the ones used in \citet{wright20} and will be described in detail in van den Busch et al.~(in preparation) for KV450. Here, we further describe how the mock catalogues are designed to resemble the DES-Y1 data. It is important to stress that this exercise is not meant to produce fully realistic mock catalogues that resemble the data in all aspects. Rather, it is aimed at producing mock catalogues that are \emph{similarly complex} as the data. As such, the mock catalogues can be used to inform us about the expected size of systematic uncertainties in the DIR calibration.

We first estimate the observed size and shape of each simulated galaxy by taking the semi major and minor axes reported in the MICE2 catalogue and adding the seeing \citep[from][]{dw18} in quadrature. Together with the 10$\sigma$ limiting magnitudes quoted in \citet{dw18} we estimate the noise level of the evolution-corrected model magnitudes. Subsequently, drawing from the corresponding Gaussian distributions, we create a noise realization for each galaxy in each band and re-calculate the magnitude uncertainty based on this realization. This yields a catalogue of `observed' magnitudes and their errors. We found that treating the limiting magnitudes from \citet{dw18} as 10$\sigma$ limits in this way results in a mock catalogue that is too shallow compared to the data, which might be attributed to aperture effects. Deliberately adapting the limits to 12$\sigma$ yields a good match between data and mocks in terms of the noise level in the four DES bands. We note that we include weak lensing magnification in all magnitude estimates although this particular aspect has virtually no impact on the results.

Subsequently, we match each mock galaxy to its nearest neighbour in the data catalogue in 4-dimensional magnitude space and assign it the responsivity weight of that galaxy in the data. This yields a properly weighted mock source sample. We run BPZ to estimate photo-$z$ for the mock galaxies using the setup described in \citet{Hildebrandt19}, but restricting the redshift range to that of MICE2 ($0.06<z<1.4$).\footnote{We also run a setup with a wider fitting range of $0<z<7$ and do not find any significant changes in the results.} This setup differs slightly from what is done in DES-Y1 \citep{Hoyle18} but the properties of the resulting photo-$z$ (scatter, bias, and outlier rate as a function of photo-$z$ and magnitude) are very similar to what is seen when comparing the DES 4-band photo-$z$ to the combined spec-$z$ sample on the data.

The next crucial step is to select samples from the mock catalogue that resemble the spec-$z$ samples used in the DIR analysis presented in Sect.~\ref{specsec}. Here, we apply the same selection criteria as the zCOSMOS \citep[$i<22.5$;][]{lilly07}, VVDS \citep[$i<24$;][]{lefevre05,vvds13}, and DEEP2 \citep[$R<24.1$ plus color selection;][]{deep13} teams to areas that correspond to the areas sampled by the data. Moreover, we implement the magnitude- and partly also redshift-dependent spectroscopic success rates reported in those papers. Where necessary, we further downsample the catalogues to yield numbers comparable to the data. This is required because the number density as a function of redshift is not identical in the simulation and the real Universe.

We find that the fiducial DEEP2 color selection yields a redshift distribution that looks somewhat different from the one in the data. This is probably due to the fact that galaxy colors in MICE2 are not fully realistic, especially at high redshift. Inspecting the location of those high-$z$ galaxies that are supposed to be targeted by DEEP2 in $B-R$ vs. $R-I$ color space, we slightly adapt those criteria to take the slightly different colors of MICE2 galaxies into account. This yields a better match to the observed spectroscopic redshift distribution. In the end, this adaptation does not have a strong influence on the results as we find by running tests with the original as well as the adapted cuts.

We select tomographic bins from the MICE2 realization of the DES-Y1 data and calibrate those with the DIR method using the mock spec-$z$ samples described above. Comparing the true mean redshifts of the galaxies in those four tomographic bins to the ones estimated from DIR on the mocks yields offsets of $\Delta z_1=\langle z_1 \rangle_{\mathrm {True}} - \langle z_1 \rangle_{\mathrm{DIR}} = -0.026$, $\Delta z_2 = -0.021$, $\Delta z_3 = -0.033$, and $\Delta z_4 = -0.012$ (see also Table~\ref{taba1label}). The exact values depend somewhat on the exact definition of the mock spec-$z$ sample. 

These results indicate that we might overestimate the true redshifts of the tomographic bins and hence underestimate $S_8$, an effect opposite - albeit smaller - to the one seen when replacing the original DES $n(z)$ with our spectroscopic re-calibration. This could be attributed to the color pre-selection of DEEP2 in combination with a magnitude space that is limited to four dimensions, such that the photometric information from the DES $griz$ filters alone is not capable of accurately breaking color-redshift degeneracies and down-weighting the high-$z$ DEEP2 galaxies. This problem was already noted in \citet{gruen17} using a similar technique. While MICE2 shows an impressive similarity to the real Universe, there is certainly the caveat that the simulation is limited to $z<1.4$. The modeling of high-$z$ tails is therefore not possible. The issue with the colors of high-$z$ galaxies further illustrates the limitations of such a mock. Whether these results hold with a mock catalogue extending further in redshift remains to be seen and will be investigated in future work.\footnote{See also \citet{hartley20} who conduct a similar simulated analysis with a different calibration sample.}

\begin{table}
\setlength\tabcolsep{18pt}
  \caption{\label{taba1label} KV450 and DES-Y1 changes in the mean redshift for each tomographic bin informed by the MICE2 mock catalogues (i.e.~${\rm Truth\,-\,{\rm DIR}_{\rm MICE2}}$). We note that these MICE2 uncertainties have conservatively been multiplied by a factor of two to account for the inherent limitations of the mock catalogues.}
  \begin{tabular}{ccc}
    \hline
    \hline
    Tom. & $\phantom{-}${KV450} & $\phantom{-}${DES-Y1}\\
          bin  & $\phantom{-}\Delta<z>$ & $\phantom{-}\Delta<z>$ \\
    \hline
    1 & $-0.048 \pm 0.010$  & $-0.026 \pm 0.016$\\
    2 & $-0.026 \pm 0.008$  & $-0.021 \pm 0.014$\\
    3 & $-0.033 \pm 0.012$  & $-0.033 \pm 0.010$\\
    4 & $\phantom{+}0.005 \pm 0.008$  & $-0.012 \pm 0.012$\\
    5 & $\phantom{+}0.013 \pm 0.008$  & $\phantom{-}$---\\
    \hline
  \end{tabular}
\end{table}

The DES-Y1 biases are comparable in size to what was found on very similar mocks resembling the KV450 data, as reported in \citet{wright20} and Table~\ref{taba1label}. While the DIR calibration on 9-band KV450 data should be less prone to systematic uncertainties than the one on the 4-band DES-Y1 data, we suppose that the greater depth of KV450 complicates the calibration and leads to biases of the same order. Despite these limitations, the biases found in the mock analysis give an indication of the systematic error inherent in the DIR calibration with typical spectroscopic catalogues. In order to address this concern, we run another cosmology parameter analysis where we apply these shifts by centering the priors on the $\Delta z_i$ parameters on these values instead of zero. As an uncertainty we use the standard deviation over 100 lines-of-sight in the mocks but multiply this by an arbitrary factor of two to account for limitations in the simulation.

\begin{figure}
\vspace{-0.86em}
\hspace{-0.43em}
\includegraphics[width=1.04\hsize]{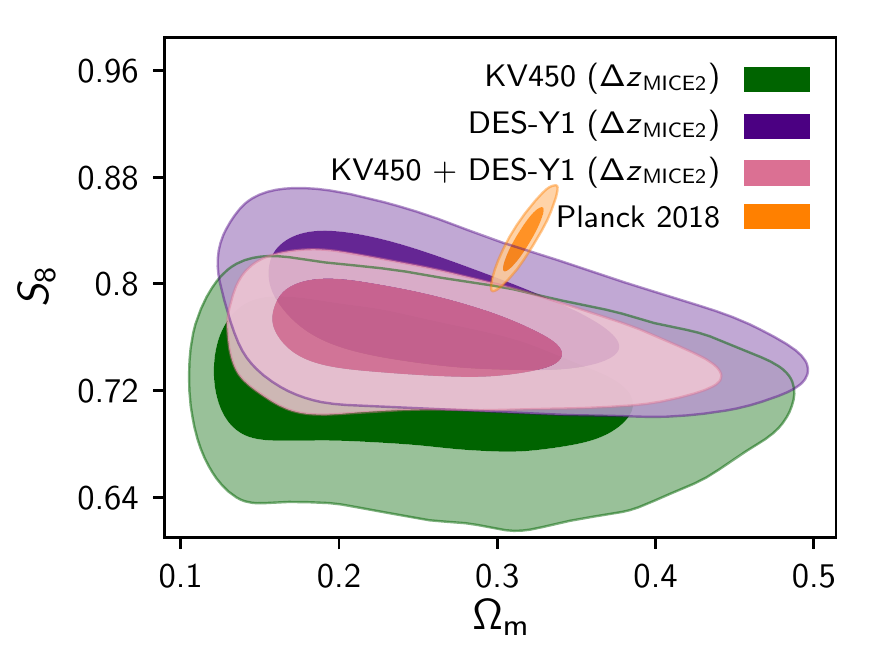}
\vspace{-2.2em}
\caption{\label{figa1label} Marginalized posterior contours in the $S_8$ -- $\Omega_{\mathrm m}$ plane (inner 68\%~CL, outer 95\%~CL) following an alternative analysis of the cosmic shear datasets with MICE2 priors on the $\Delta z_i$ parameters. We show KV450 in green, DES-Y1 in purple, KV450 + DES-Y1 in pink, and Planck 2018 in orange.}
\end{figure}

Following this approach, we find that $S_8 = 0.721^{+0.045}_{-0.032}$ for KV450, $S_8 = 0.777^{+0.035}_{-0.033}$ for DES-Y1, and $S_8 = 0.760^{+0.026}_{-0.023}$ for KV450 + DES-Y1 (as shown in Fig.~\ref{figa1label}). These constraints correspond to changes of $\Delta S_8 = [-0.014, +0.012, -0.002]$ and $\Delta\chi^2 = [0.52, 2.0, 2.8]$\footnote{These changes in the goodness of fit are obtained for data vectors of $\{195, 227, 422\}$ elements in size and thereby \{$180$, $211$, $397$\} degrees of freedom for KV450, DES-Y1, and KV450 + DES-Y1, respectively.}  relative to our fiducial results in the main body of the paper for KV450, DES-Y1, and KV450 + DES-Y1, respectively. Given the comparable size of the applied MICE2 and bootstrap errors on the $\Delta z_i$ parameters, we do not find significant differences in the size of the $S_8$ uncertainties (the largest difference corresponds to a 2\% increase in the uncertainty).

We note that the KV450 constraint on $S_8$ shifts towards lower values despite substantial negative $\Delta z_i$ shifts in the first three tomographic bins. This is explained by the greater constraining power of the higher redshift fourth and fifth bins which exhibit positive shifts in their mean redshifts. The change in $S_8$ is $-0.017$ for DES-Y1 relative to the COSMOS-2015 calibrated redshift distributions, which corresponds to a $0.5\sigma$ shift in terms of the larger DES-Y1 uncertainty in the KV450 setup (as compared to the fiducial shift of $0.8\sigma$; noting the significance of both shifts increase in terms of the original DES-Y1 uncertainty). In other words, the MICE2 mocks suggest that the redshift distributions from the pre-revised DIR in the case of KV450 and from COSMOS-2015 in the case of DES-Y1 both result in an overestimated posterior mean of $S_8$ (by $0.014$ and $0.017$ respectively). Here, the concordance in the MICE2-revised $S_8$ constraints of KV450 and DES-Y1 is at the $1.1\sigma$ level (as compared to the stronger fiducial concordance in $S_8$ of $0.6\sigma$), while the combined KV450 + DES-Y1 constraint on $S_8$ remains unchanged at $0.1\sigma$ relative to the fiducial result.

In summary, this analysis on the MICE2 mocks illustrates the importance of realistic mock catalogues for future analyses of weak lensing surveys. In particular, a very wide redshift range is desirable to properly account for photo-$z$ outliers in the systematic error estimates.

\section{Further acknowledgements}
\label{moreacc}

Part of this work was performed using the DiRAC Data Intensive service at Leicester operated by the University of Leicester IT Services, and DiRAC@Durham managed by the Institute for Computational Cosmology, which form part of the STFC DiRAC HPC Facility (\diracaddress) acknowledging BEIS and STFC grants STK0003731, STR0023631, STR0010141, STP0022931, STR0023711, STR0008321. 
We acknowledge support from the European Research Council under grant numbers 693024 (SJ, DT), 770935 (HHi, AHW), 647112 (CH, MA, TT). SJ and DT acknowledge support from the Beecroft Trust and STFC. HHi is supported by Emmy Noether (Hi 1495/2-1) and Heisenberg grants (Hi 1495/5-1) of the Deutsche Forschungsgemeinschaft. NEC is supported by a Royal Astronomical Society research fellowship. HHo and AK acknowledge support from Vici grant 639.043.512, financed by the Netherlands Organisation for Scientific Research (NWO). KK acknowledges support by the Alexander von Humboldt Foundation. LM acknowledges support from STFC grant ST/N000919/1. TT acknowledges funding from the European Union's Horizon 2020 research and innovation program under the Marie Sk{l}odowska-Curie grant agreement No 797794.
We are indebted to the staff at ESO-Garching and ESO-Paranal for managing the observations at VST and VISTA that yielded the data presented here. Based on observations made with ESO Telescopes at the La Silla Paranal Observatory under programme IDs 177.A-3016, 177.A-3017, 177.A-3018, 179.A- 2004, 298.A-5015, and on data products produced by the KiDS consortium. 
{This project used public archival data from the Dark Energy Survey (DES). Funding for the DES Projects has been provided by the U.S. Department of Energy, the U.S. National Science Foundation, the Ministry of Science and Education of Spain, the Science and Technology Facilities Council of the United Kingdom, the Higher Education Funding Council for England, the National Center for Supercomputing Applications at the University of Illinois at Urbana-Champaign, the Kavli Institute of Cosmological Physics at the University of Chicago, the Center for Cosmology and Astro-Particle Physics at the Ohio State University, the Mitchell Institute for Fundamental Physics and Astronomy at Texas A\&M University, Financiadora de Estudos e Projetos, Funda{\c c}{\~a}o Carlos Chagas Filho de Amparo {\`a} Pesquisa do Estado do Rio de Janeiro, Conselho Nacional de Desenvolvimento Cient{\'i}fico e Tecnol{\'o}gico and the Minist{\'e}rio da Ci{\^e}ncia, Tecnologia e Inova{\c c}{\~a}o, the Deutsche Forschungsgemeinschaft, and the Collaborating Institutions in the Dark Energy Survey.
The Collaborating Institutions are Argonne National Laboratory, the University of California at Santa Cruz, the University of Cambridge, Centro de Investigaciones Energ{\'e}ticas, Medioambientales y Tecnol{\'o}gicas-Madrid, the University of Chicago, University College London, the DES-Brazil Consortium, the University of Edinburgh, the Eidgen{\"o}ssische Technische Hochschule (ETH) Z{\"u}rich,  Fermi National Accelerator Laboratory, the University of Illinois at Urbana-Champaign, the Institut de Ci{\`e}ncies de l'Espai (IEEC/CSIC), the Institut de F{\'i}sica d'Altes Energies, Lawrence Berkeley National Laboratory, the Ludwig-Maximilians Universit{\"a}t M{\"u}nchen and the associated Excellence Cluster Universe, the University of Michigan, the National Optical Astronomy Observatory, the University of Nottingham, The Ohio State University, the OzDES Membership Consortium, the University of Pennsylvania, the University of Portsmouth, SLAC National Accelerator Laboratory, Stanford University, the University of Sussex, and Texas A\&M University.
Based in part on observations at Cerro Tololo Inter-American Observatory, National Optical Astronomy Observatory, which is operated by the Association of Universities for Research in Astronomy (AURA) under a cooperative agreement with the National Science Foundation.}

\end{appendix}

\end{document}